\documentclass[twocolumn]{aastex62}

\usepackage{graphicx}

\newcommand{\UDA}{Instituto de Astronom\'ia y Ciencias Planetarias, Universidad de Atacama, Copayapu 485, Copiap\'o, Chile}
\newcommand{\UNAB}{Depto. de Cs. F\'isicas, Facultad de Ciencias Exactas, Universidad Andr\'es Bello, Av. Fern\'andez Concha 700, Las Condes, Santiago, Chile}
\newcommand{\VATICAN}{Vatican Observatory, V-00120 Vatican City State, Italy}
\newcommand{\DAME}{Department of Physics and JINA Center for the Evolution of the Elements, University of Notre Dame, Notre Dame, IN 46556, USA}
\newcommand{\SAO}{Universidade de S\~ao Paulo, IAG, Rua do Mat\~ao 1226, Cidade Universit\'aria, S\~ao Paulo 05508-900, Brazil}    
\newcommand{\AIP}{Leibniz-Institut f\"ur Astrophysik Potsdam (AIP), An der Sternwarte 16, 14482 Potsdam, Germany}
\newcommand{\RIO}{Laborat\'orio Interinstitucional de e-Astronomia - LIneA, Rua Gal. Jos\'e Cristino 77, Rio de Janeiro, RJ - 20921-400, Brazil}
\newcommand{\UDEC}{Departamento de Astronom\'\i a, Casilla 160-C, Universidad de Concepci\'on, Concepci\'on, Chile}

\newcommand{\UBO}{Centro de Investigaci\'on en Astronom\'ia, Universidad Bernardo O'Higgins, Avenida Viel 1497, Santiago, Chile}
\newcommand{\UNAM}{Instituto de Astronom\'ia, Universidad Nacional Aut\'onoma de M\'exico, Apdo. Postal 70264, M\'exico D.F., 04510, M\'exico}
\newcommand{\UTINAM}{Institut Utinam, CNRS UMR 6213, Universit\'e Bourgogne-Franche-Comt\'e, OSU THETA Franche-Comt\'e, Observatoire de Besan\c{c}on, BP 1615, 25010 Besan\c{c}on Cedex, France}
\newcommand{\APO}{Apache Point Observatory and New Mexico State University, P.O. Box 59, Sunspot, NM, 88349-0059, USA}
\newcommand{\Moscow}{Sternberg Astronomical Institute, Moscow State University, Moscow}
\newcommand{\UCN}{Instituto de Astronom\'ia, Universidad Cat\'olica del Norte, Av. Angamos 0610, Antofagasta, Chile}
\newcommand{\VIRGINIA}{Department of Astronomy, University of Virginia, Charlottesville, VA 22904, USA}

\newcommand{\UNAMB}{Instituto de Radioastronom\'ia y Astrof\'isica, Universidad Nacional Aut\'onoma de M\'exico, Apdo.postal 3-72, Morelia Michoac\'an, 58089, Mexico}

\newcommand{\COL}{Astronomical Observatory, Universidad de Nari\~no, Sede VIIS, Avenida Panamericana, Pasto, Nari\~no, Colombia}
\newcommand{\COLL}{Departamento de Fisica de la Universidad de Nari\~no, Torobajo Calle 18 Carrera 50, Pasto, Narino, Colombia}

\begin{document}

\title{APOGEE-2 Discovery of a Large Population of Relatively High-Metallicity Globular Cluster Debris}

\correspondingauthor{Jos\'e G. Fern\'andez-Trincado}
\email{jfernandezt87@gmail.com}

\author[0000-0003-3526-5052]{Jos\'e G. Fern\'andez-Trincado}
\affil{\UCN}
\affil{\UDA}

\author[0000-0003-4573-6233]{Timothy C. Beers}
\affil{\DAME}

\author{Anna. B. A. Queiroz}
\affil{\AIP}	

\author{Cristina Chiappini}
\affil{\AIP}	
\affil{\RIO}	

\author[0000-0002-7064-099X]{Dante Minniti}
\affil{\UNAB}
\affil{\VATICAN}

\author[0000-0001-9264-4417]{Beatriz Barbuy}
\affil{\SAO}

\author[0000-0003-2025-3147]{Steven R. Majewski}
\affil{\VIRGINIA}

\author{Mario Ortigoza-Urdaneta}
\affil{\UDA}

\author{Christian Moni Bidin}
\affil{\UCN}

\author{Annie C. Robin}
\affil{\UTINAM}	

\author{Edmundo Moreno}
\affil{\UNAM}

\author{Leonardo Chaves-Velasquez}
\affil{\UNAMB}
\affil{\COL}
\affil{\COLL}

\author{Sandro Villanova}
\affil{\UDEC}

\author{Richard R. Lane}
\affil{\UBO}

\author{Kaike Pan}
\affil{\APO}

\author{Dmitry Bizyaev}
\affil{\APO}
\affil{\Moscow}

\begin{abstract}
We report the discovery of a new, chemically distinct population of relatively high-metallicity ([Fe/H] $> -0.7$) red giant stars with super-solar [N/Fe] ($\gtrsim +0.75$) identified within the bulge, disk, and halo of the Milky Way. This sample of stars was observed during the second phase of the Apache Point Observatory Galactic Evolution Experiment (APOGEE-2); the spectra of these stars are part of the seventeenth Data Release (DR~17) of the Sloan Digital Sky Survey. We hypothesize that this newly identified population was formed in a variety of progenitors, and are likely made up of either fully or partially destroyed metal-rich globular clusters, which we refer to as Globular Cluster Debris (GCD), identified by their unusual photospheric nitrogen abundances. It is likely that some of the GCD stars were probable members of the \textit{Gaia}-Enceladus-Sausage accretion event, along with clusters formed in situ.
\end{abstract}

\section{INTRODUCTION} 
\label{section1}

It is well known that the Milky Way (MW) is populated by a great variety ($\sim$170) of ancient globular clusters (GCs)--\citep[][and references therein]{Gratton2004, Vasiliev2021, Baumgardt2021}, and perhaps a hundred new low-luminosity candidates uncovered in the inner Galaxy \citep{Minniti2017}. 

The MW GCs measured so far span a wide range of metallicities, above the so-called ``metallicity floor" at [Fe/H]$=-2.5$, with only two cases that have metallicities close to this value, VVV~CL001 at [Fe/H]$=-2.45$ \citep[][]{Fernandez-Trincado2021} and ESO280$-$SC06  at [Fe/H]$=-2.48$ \citep[][]{Simpson2018}. 

While many of these systems have long survived, a handful of unique field stars with chemical patterns differing from the typical patterns observed in the MW and other Local Group galaxies suggest that many of the surviving and destroyed GCs have deposited part or all of their stellar content into the bulge, disk, and halo of the MW \citep[see, e.g.,][]{Nissen2010, Fernandez-Trincado2016, Fernandez-Trincado2017, Fernandez-Trincado2019b, Hanke2020, Wan2020}.

On the metal-rich end, a handful of GCs have been found to exhibit metallicities very near solar metallicity, such as Pal~10 \citep{Harris2010}. Motivated by the prospects of unearthing new examples of such GC debris, we have searched for the presence of  relatively high-metallicity ([Fe/H]$\gtrsim-0.7$) GC debris throughout the MW within the footprint of the APOGEE-2 survey.

In this Letter, we report the discovery of a significant population of high-metallicity Globular Cluster Debris (GCD) stars within the bulge, disk, and halo of the MW likely associated with small pieces of destroyed GCs and identified by their unusual nitrogen abundances.

\begin{figure*}
	\begin{center}
		\includegraphics[width=180mm]{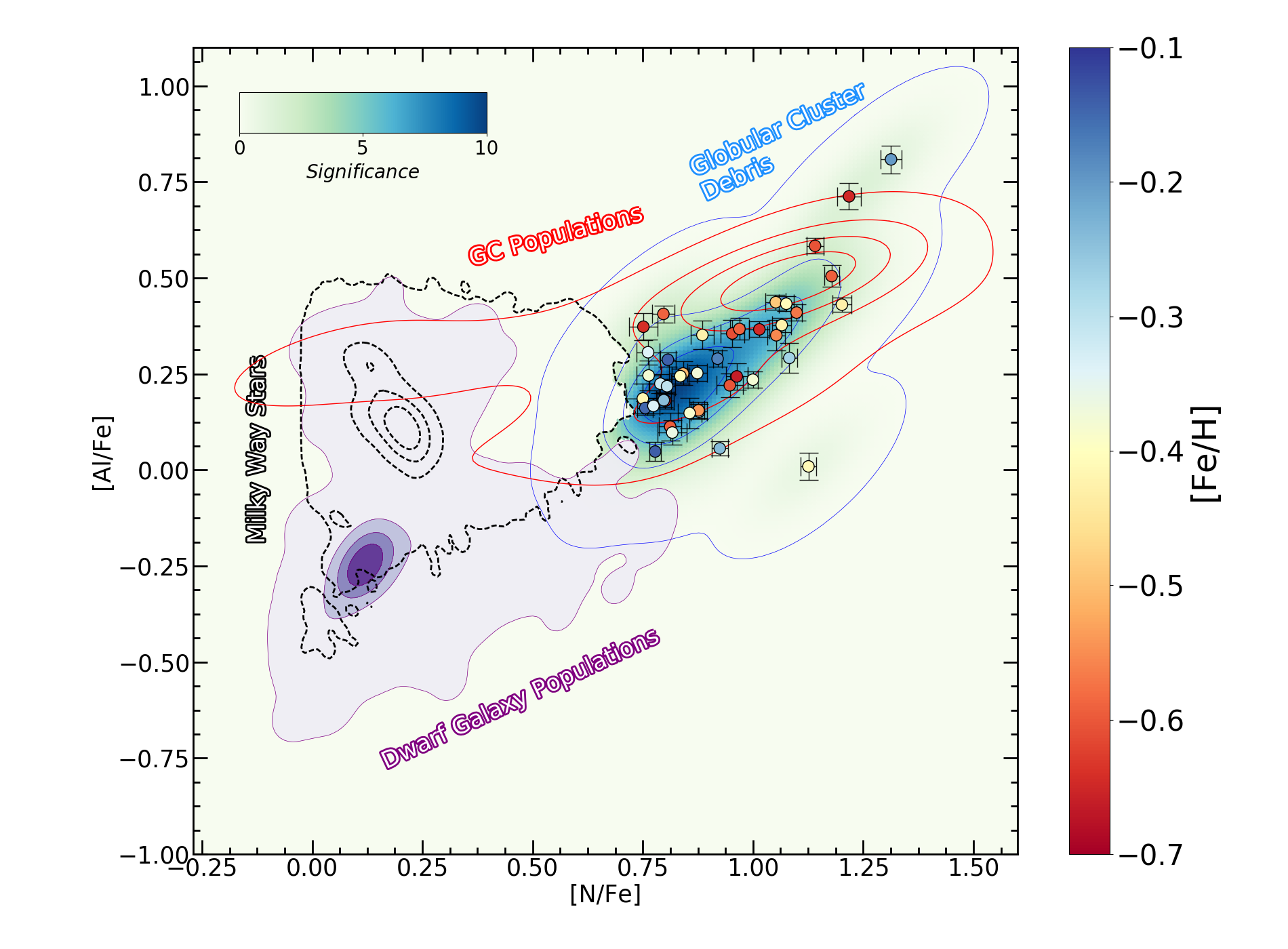}
		\caption{Distribution of [Al/Fe] vs. [N/Fe] for all stars used in our analysis, which all have been chosen to be of high metalllicity ([Fe/H] $> -0.7$). Also included are KDE models and contours showing the density of objects belonging to the MW Stars (black contours), Dwarf Galaxy Populations (purple contours), GC Populations (red contours), and the newly identified population--the GCD stars (blue contours and circle symbols). Stars among the GCD are also overlaid with dot symbols with color-coding based on their [Fe/H]. The background green to blue color highlighted by the inset bar indicate the significance of the GCD stars as compared to the background density level.}
		\label{Figure1}
	\end{center}
\end{figure*}

\section{DATA} 
\label{section2}

We employ interim data from the seventeenth data release (DR~17) of the second generation of the Apache Point Observatory Galactic Evolution Experiment (APOGEE-2) survey \citep{Majewski2017}, which is one of the programs of the Sloan Digital Sky Survey \citep[SDSS-IV;][]{Blanton2017}. The APOGEE instruments are high-resolution ($R \sim 22,500$), near-infrared (collecting $\sim$2/3 of the \textit{H}-band: $15145$--$16960$ \AA{}; vacuum wavelengths) spectrographs \citep{Wilson2019} that operate on the Sloan 2.5m telescope \citep{Gunn2006} at Apache Point Observatory (APOGEE-2N) and on the Ir\'en\'ee du Pont 2.5m telescope \citep{Bowen1973} at Las Campanas Observatory (APOGEE-2S). 

As of January 2021, the dual APOGEE instruments have observed more than 700,000 stars across the MW. \citet{Zasowski2017}, \citet{Beaton2021}, and \citet{Santana2021} provide a detailed overview of the targeting strategy of the APOGEE-2 survey. Spectra are reduced as described in \citet[][]{Nidever2015}, and analyzed using the APOGEE Stellar Parameters and Chemical Abundance Pipeline \citep[][hereafter \texttt{ASPCAP}/APOGEE-2]{Garcia2016} and the libraries of synthetic spectra described in \citet{Zamora2015}. The accuracy and precision of the atmospheric parameters and chemical abundances are extensively analyzed in \citet{Holtzman2018}, while details regarding the customized \textit{H}-band line list are fully described in \citet{Shetrone2015} and \citet{Smith2021}. 

\begin{figure}
	\begin{center}
		\includegraphics[width=80mm]{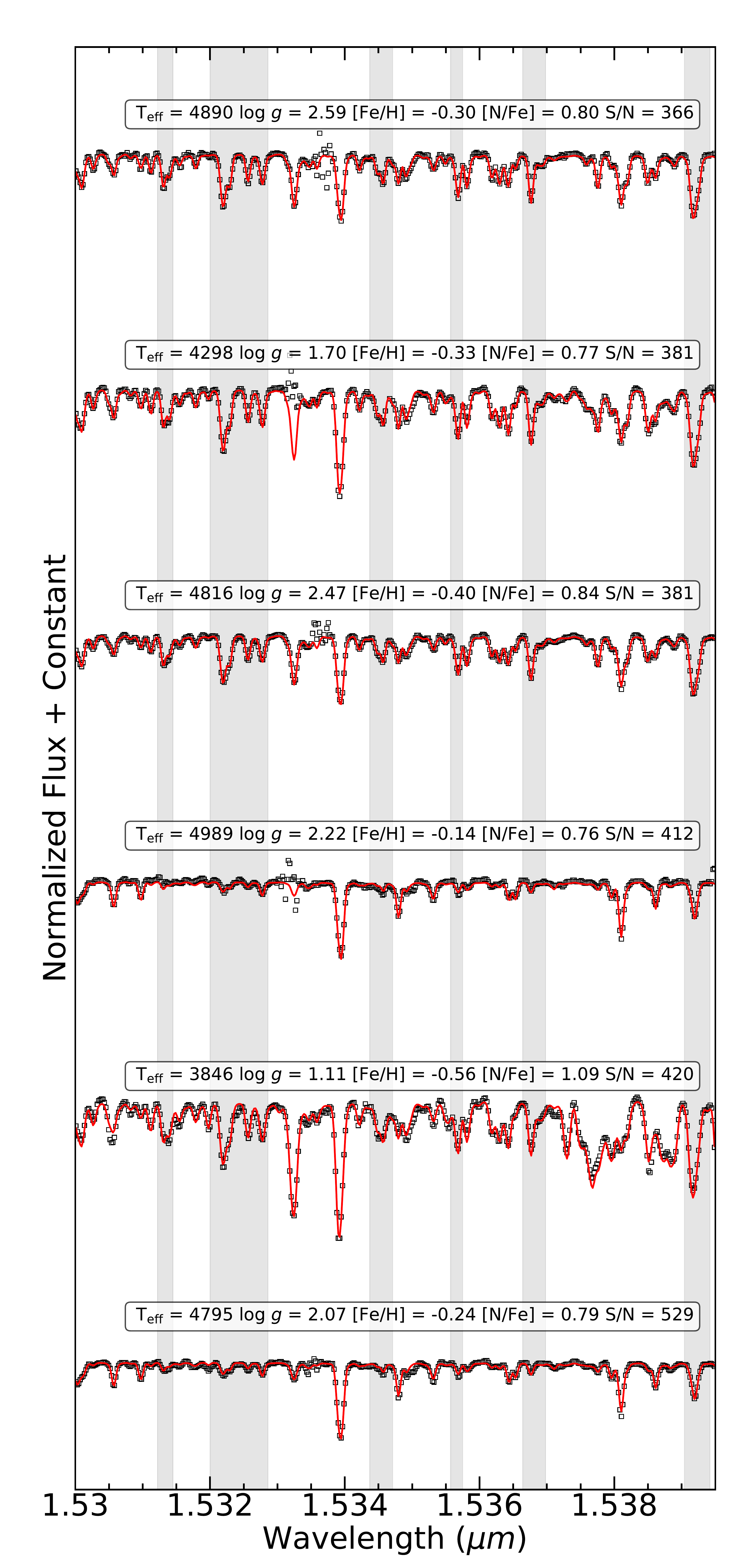}
		\caption{High-resolution near-IR \textit{H}-band spectra (black empty squares) of six arbitrary stars identified as potential members of the GCD population, covering a small portion of the observed spectral regions around the $^{12}$C$^{14}$N band (grey bands). The best-fit models based on the \texttt{ASPCAP}/APOGEE-2 spectral synthesis are superposed (red lines).}
		\label{Figure2}
	\end{center}
\end{figure}

\section{SAMPLE SELECTION} 
\label{section3}

We restrict our sample to giant stars ($\log$ \textit{g} $<3.6$) that have been flagged as \texttt{ASPCAPFLAG}$=0$, AL\_FE\_FLAG$=0$, N\_FE\_FLAG$=0$, and have a spectral signal-to-noise ratio (SNR) larger than 70. Our sample is also restricted to 3200\,K $< T_{\rm eff} <$ 5500\,K, in order to avoid large uncertainties in [N/Fe] due the weakness of the $^{12}$C$^{14}$N, $^{12}$C$^{16}$O, and $^{16}$OH molecules, which together are required to determine the nitrogen, oxygen, and carbon abundances. Moreover, because we are interested in searching for relatively high-metallicity ($-0.7<$ [Fe/H] $\lesssim-0.1$) stellar debris with possible GC origin, we restrict our analysis to relatively carbon-poor stars, [C/Fe] $\lesssim+0.15$ \citep[see][for a discussion]{Martell2016, Schiavon2017, Fernandez-Trincado2019a, Fernandez-Trincado2020b, Fernandez-Trincado2020a, Fernandez-Trincad2020d}, because such stars are typically found in GCs \citep[see e.g.,][]{Meszaros2020}, and at the same time minimize potential contamination by objects such as CH stars \citep{Karinkuzhi2015}, whose surface abundances may have been modified by mass-transfer from low-mass companion AGB stars \citep{Karinkuzhi2015}. The cuts above return a data set of 143,141 unique sources, after removing duplicate entries. This sample have been divided in several main populations as described in next section.

\section{Potential High-metallicity Globular Cluster Debris} 
\label{section4}

As nitrogen enrichment in GCs is even larger than what is possible/expected from pure mixing processes \citep[see, e.g.,][]{Shetrone2019} occurring in non-cluster stars, we start by evaluating the chemical species (N and Al) that typically participate in the characteristic GC abundance patterns over a wide range of metallicities \citep[see, e.g.,][]{Meszaros2020}. In the following, we describe our two-step selection of outliers in the [N/Fe]--[Al/Fe] space.

First, we carried out a population analysis in the [Al/Fe]--[N/Fe] plane using the \textit{k}-means clustering approach, as described in \citet{Ivezic2014}. The \textit{k}-means clustering algorithm revealed the existence of  a significant population of stars that is separated relatively cleanly from the MW field stars approximately between \{[N/Fe],[Al/Fe]\}$=$\{$+0.75$, $-0.05$\} and \{$+1.55$, $+1.1$\}, and have been found to be $>$ 4$\sigma$ above the typical MW field stars in the [Fe/H]--[N/Fe] plane at fixed metallicity. After removing 50 well-studied APOGEE-2 GC stars \citep{Meszaros2020}, and $\sim$145  potential GCs candidates\footnote{ These sources were not included in our analysis, as the potential progenitors are primarily located in regions heavily reddened which requires a  more exhaustive treatment of their atmospheric parameters and elemental abundances.} not analyzed in \citet{Meszaros2020}. Thus, the final sample was reduced to a total of 42 unique stars, marked with circles in Figure \ref{Figure1}. There are no known GCs within an angular separation of 0.5$^{\circ}$ from these stars.  

For comparison, with our relatively high-metallicity outliers (potential GCDs), we define and include in Figure \ref{Figure1} four main populations of similar metallicity ([Fe/H]$>-$0.7) stars: MW stars (140,270 source in the thin an thick disk as well as likely metal-rich halo stars),  confirmed dwarf galaxy stars (2,634 sources in the Sagittarius, and the Large and Small Magellanic Clouds) from \citet{Helmi2018} with \texttt{ASPCAP}/APOGEE-2 DR~17 abundances (dwarf galaxy populations), and relatively high-metallicity GC stars (50 sources) from \citet{Meszaros2020} having \texttt{ASPCAP}-measured abundances in DR17. We ran a Kernel Density Estimation (KDE) model over the stars in every population, as shown in Figure \ref{Figure1}. As can be appreciated from inspection of this Figure \ref{Figure1}, the MW stars and dwarf galaxy populations lie below [N/Fe] $\lesssim+0.75$ over a wide range of [Al/Fe] abundance ratios, and well below [Al/Fe] $\lesssim +0.5$. In contrast, the relatively high-metallicity GCs span a range in [N/Fe], which is much wider than that of the MW and dwarf galaxy populations, and exhibits [Al/Fe] abundance ratios from the solar level to [Al/Fe] $< +1.0$.

Figure \ref{Figure1} also shows some main characteristics of the newly identified GCD population. For instance, Figure \ref{Figure1} reveal that the GCD clearly occupies the locus dominated by the nitrogen enriched ([N/Fe]$\gtrsim+0.7$) GC populations, suggesting that GCD stars may have formed in different sites than typical MW stars, unless they were part of systems with contribution from ``spinstars" \citep{Frischknecht2016}, which would alternatively explain the anomalous abundance signature. Close inspection of Figure \ref{Figure1} also reveals a fairly clear clump of potential GCD stars that are not located in the main bulk of the KDE of the MW stars or dwarf galaxy populations, exceeding the background level by a factor of more than 10. This suggests the existence of a statistically significant newly discovered stellar population of potential GCD stars in the relatively high-metallicity regime. 

The [N,Al/Fe]-peak is clearly visible at ([N/Fe],[Al/Fe])$\gtrsim$ ($+$0.75,$+$0.0) in Figure \ref{Figure1}; a set of contour lines is provided as a visual aid. The statistical significance of the detection of newly identified GCD stars confirms and reinforces the existence of a newly discovered, relatively high-metallicity stellar sub-population whose origin is different from that that of MW stars, and clearly well-separated from the MW and dwarf galaxy stellar systems. 

Figure \ref{Figure2} shows examples of the typical high-S/N APOGEE-2 DR~17 spectra of  arbitrarily selected stars belonging to the GCD population, covering a small portion of the observed spectral regions around the remarkably strong $^{12}$C$^{14}$N lines, along with corresponding best-fit models from the \texttt{ASPCAP}/APOGEE-2 spectral synthesis. It is clear that the \textit{ASPCAP} analysis is performing well on these stars, in particular, in the regions of the cyanogen bands.

\begin{figure*}
	\begin{center}
		\includegraphics[width=190mm]{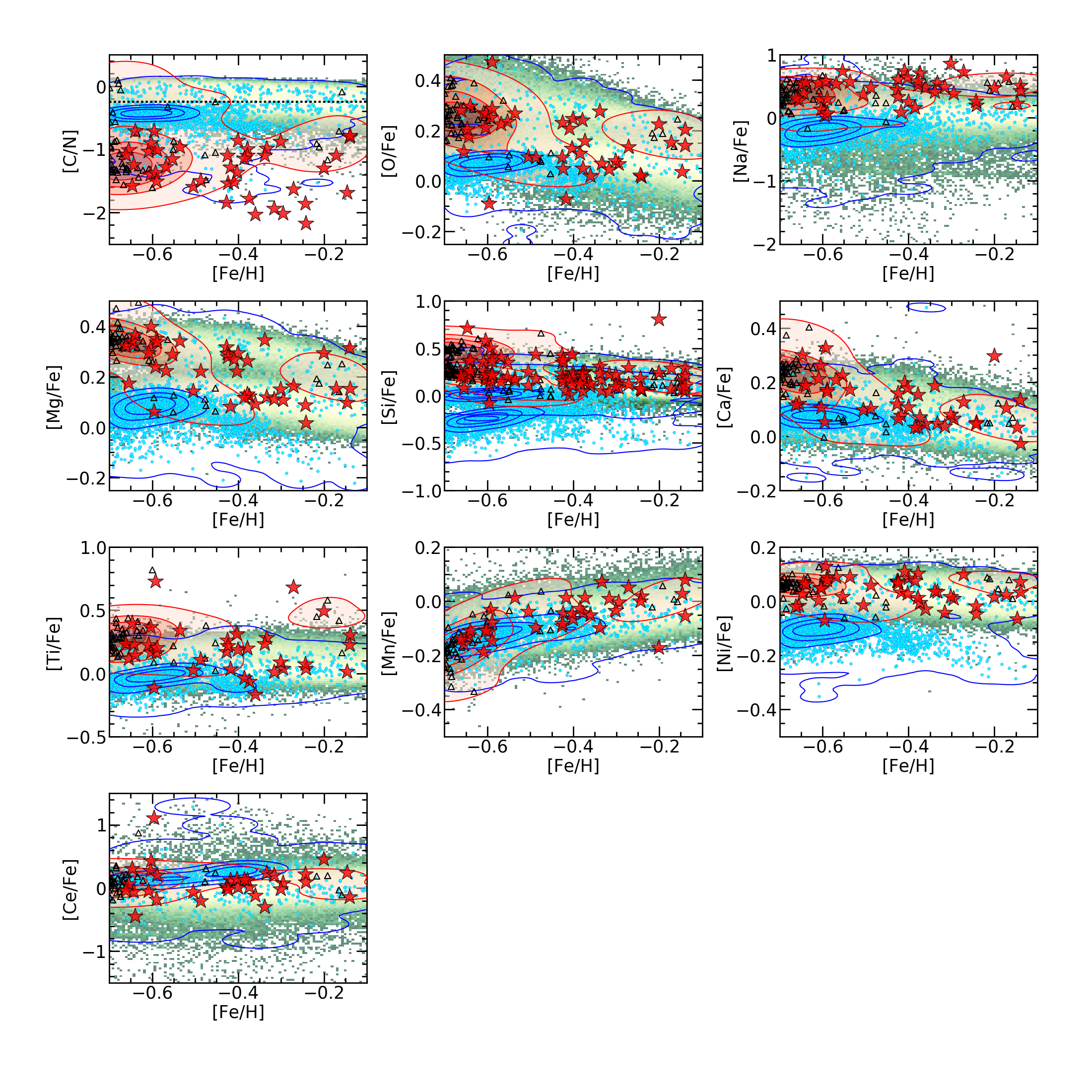}
		\caption{Chemical-abundance patterns of selected elements for the relatively high-metallicity GCD population (red filled stars) compared to the MW stars (green 2D density histogram), dwarf galaxy populations (cyan circles and blue contours), and GC populations (black empty triangles and red contours). The black dotted line highlighted in the [Fe/H]--[C/N] plane indicates the limit above which extra-mixing episodes are expected to take place \citep[see, e.g.,][]{Shetrone2019}.}
		\label{Figure3}
	\end{center}
\end{figure*}

Figure \ref{Figure3} shows the chemical-abundance patterns of selected APOGEE-2 DR~17 chemical species for the relatively high-metallicity GCD stars compared to those for the MW stars, dwarf galaxy populations, and GC populations. While the GCD stars largely follow the expected chemical enrichment of the relatively high-metallicity GC populations, we cannot rule out possible contamination from the incidence of variable stars or mass-transfer events in our sample, which could be responsible for producing the anomalous abundance variations (primarily with N and Al) observed in this relatively high-metallicity regime. 

Even though our APOGEE-2 DR~17 data do not provide any strong evidence for variability in radial velocity over the period of the APOGEE-2  DR~17 observations, we identify one star in our sample (2M06410076$-$6926199) with [N/Fe] $>+0.87$, [Al/Fe] $\sim+0.16$, and [Mg/Fe] $\sim+0.34$ that has been classified (with a 97\% of probability) as a semi-regular variable (likely a highly evolved AGB star) with a period of 17.17 days and low amplitude (0.08 mag) in the ASAS-SN Catalog \citep{Jayasinghe2021}. Unfortunately, the \textit{s}-process ([Ce/Fe]) abundance ratio for this star is not determined by the \texttt{ASPCAP}/APOGEE-2 pipeline, which makes it difficult to provide a rough estimate of the mass of the presumed AGB.

Figure \ref{Figure3} also reveals that our sample lies well below [C/N] $\sim-0.25$, which indicates that the chemical anomalies observed in our sample are unlikely to be affected by extra-mixing processes \citep[see, e.g.,][]{Shetrone2019}. This suggests that whatever process is responsible for the abundances observed in the GCD population is similar to that associated with the production of this signature for relatively high-metallicity GCs.  

\section{Dynamical Properties of Selected Globular Cluster Debris Stars} 

We study the kinematics and dynamical properties of our sample by making use of the GravPot16\_VAC\_DR17 Value Added Catalog\footnote{https://internal.sdss.org/dr17/datamodel/files/\\APOGEE\_GRAVPOT16/GravPot16\_VAC\_DR17.html} (Fern\'andez-Trincado et al., in preparation) of ensemble orbits integrated over a 3 Gyr timespan with the \texttt{GravPot16}\footnote{https://gravpot.utinam.cnrs.fr} model \citep{Fernandez-Trincado2020}. For the orbit computations we assumed a bar pattern speed of 41$\pm$10 km s$^{-1}$ kpc$^{-1}$ \citep{Sanders2019}. We note that our model has some limitations in the processes considered; e.g., secular changes in the adopted MW potential as well as dynamical friction are not included.

\begin{figure*}
	\begin{center}
		\includegraphics[width=190mm]{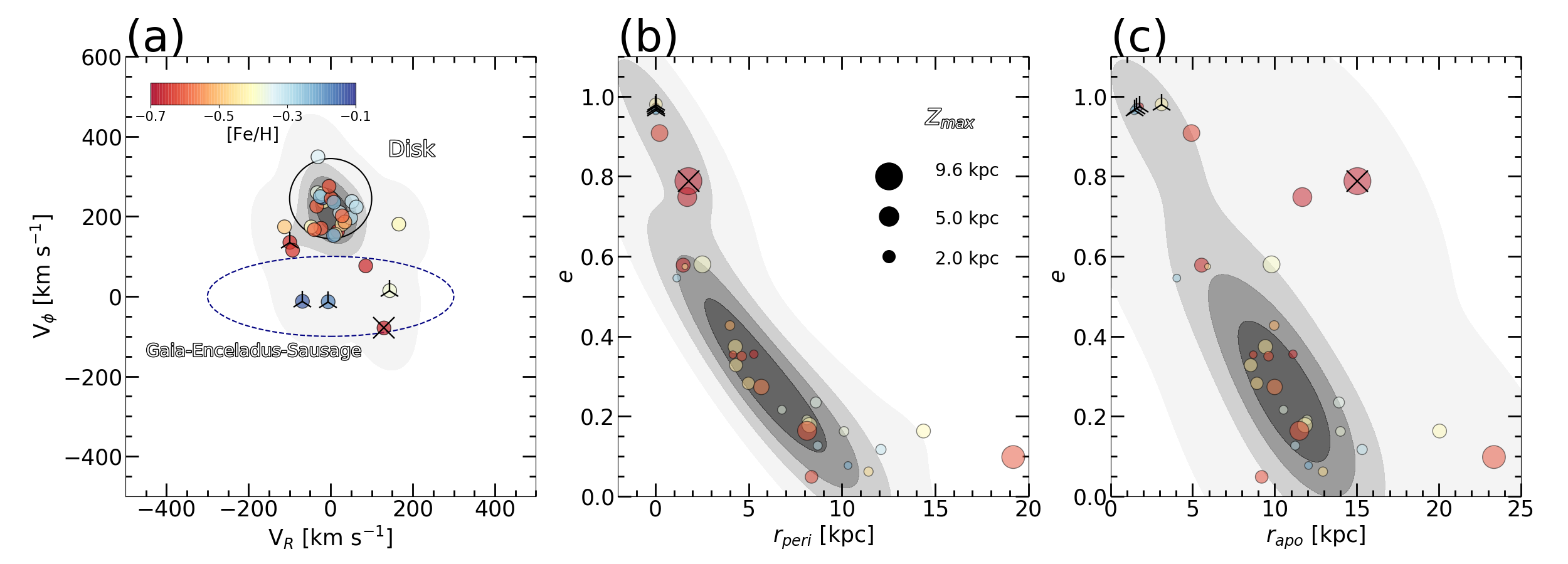}
		\caption{Kinematics and orbital-element properties of selected GCD stars. The color-coding of the symbols in the panels is based on their [Fe/H], while the gray background shading represents the Kernel Density Estimation (KDE) model of the overlaid points. The distribution of the velocity component V$_{\rm R}$ vs. V$_{\phi}$ is shown in panel (a). The blue dashed line represents the approximate region for sources associated with the \textit{Gaia}-Enceladus-Sausage in V$_{\rm R}$ vs. V$_{\phi}$, based on \citet{Belokurov2018}. The area occupied by disk-like stars is also highlighted with the black empty circle. The perigalactic and apogalactic distances, as a function of the orbital eccentricity, are shown in panels (b) and (c), respectively. The point sizes reflect their maximum vertical heights above the Galactic plane (${\rm Z_{max}}$), with decreasing size for decreasing ${\rm Z_{max}}$. Three black points at 2, 5, and 9.6 kpc are provided as a visual aid. Black up-tick symbols mark the stars with P-R orbital configurations (see text), while the ``X" symbol indicates the star with a retrograde orbit.}
		\label{Figure4}
	\end{center}
\end{figure*}

The orbit calculations were performed by adopting a simple Monte Carlo approach that considers the errors of the observables. The resulting values and their errors were taken as the 16$^{\rm th}$, 50$^{\rm th}$, and 84$^{\rm th}$ percentiles from the generated distributions. Heliocentric distances were estimated with the \texttt{StarHorse} code \citep[see, e.g.,][]{Queiroz2018, Queiroz2020, Queiroz2021}, proper motions are from \textit{Gaia} EDR~3 \citep{Brown2021}, and radial velocities are provided from the APOGEE-2 DR~17 database. Figure \ref{Figure4} shows the resulting orbital elements for the 32 out of 42 stars in our sample that have a \textit{Gaia} re-normalized unit weight error (\texttt{RUWE}) less than 1.4 (indicating the high quality of their astrometric solutions).

We find that the great majority of the stars  (27 out of 32) in our sample tagged as GCD members exhibit prograde orbits, with the exception of five stars. Among those five, four exhibit  the unusual behavior of orbits that change their sense of motion from prograde to retrograde (P$-$R orbits) during the integration time, and one star that is in retrograde motion. 

We also find that three of the stars with P$-$R orbits, the star on a retrograde orbit, and one star with a prograde orbit have kinematical properties compatible with the position of the \textit{Gaia}-Enceladus-Sausage (GES) accretion event \citep{Belokurov2018}, as can be appreciated from inspection of Figure \ref{Figure4}(a). Like GES stars, these GES candidates in our sample have radial/eccentric orbits. The stars in the P$-$R configurations are on bulge-like orbits (r$_{apo} < 3.5$ kpc) with small vertical excursions above the Galactic plane (Z$_{\rm max} < 2.5$ kpc), while the retrograde star exhibits a radial and high-eccentricity, halo-like orbit, with large vertical excursions from the Galactic plane (Z$_{\rm max} > 9$ kpc); see Figures \ref{Figure4}(b) and (c). Strikingly, three of the stars in our sample with GES-like kinematics have [Fe/H] $>-0.38$, and are strongly enriched in nitrogen ([N/Fe] $\gtrsim +0.77$), with [Al/Fe] abundance ratios ranging from $+0.05$ to $+0.29$, and sodium enrichment above [Na/Fe] $= +0.27$. Thus, the very peculiar dynamics of these stars, combined with the enrichment levels of their constituent chemical elements, make these five stars potential members of dissipated metal-rich ([Fe/H] $>-0.7$) GCs that are likely associated with the progenitor GES dwarf galaxy.

Figure \ref{Figure4}(a) also reveals that the majority of stars on prograde orbits exhibit disk-like kinematics, with typical orbital eccentricities below $e = 0.6$, and small vertical excursions below the Galactic plane (Z$_{\rm max} \lesssim 3.0$ kpc), with a few exceptions on halo-like orbits (and Z$_{\rm max} > 3.0$ kpc). The dark-gray shadow region in Figures \ref{Figure4}(b) and (c) reveals that stars  (13 out of the 32) with lower eccentricities ($e \lesssim$0.4) exhibit a peak in their perigalactocentric distances between 4 and 9 kpc, apogalactocentric distances between $\sim$8.5 and 12 kpc, and are on in-plane orbits with Z$_{\rm max} < 0.5$ kpc; this places these stars on disk-like orbits very near the Solar Neighborhood, with radial excursions reaching the co-rotation radius. The stars with prograde orbits outside these dynamical limits are likely on inner-halo-like orbits. 

Our dynamical analysis clearly reveals that there are at least four different dynamical families of stars represented within the GCD sample and suggesting a variety of origin scenarios: (\textit{i}) the group of stars that lives in bulge-like orbits likely ejected from in situ disk/bulge GCs; \textit{ii}) the stars dynamically lost to the inner-most parts of the MW,  from a massive accreted system, such as the GES dwarf galaxy; (\textit{iii}) stars in the Solar Neighborhood with in-plane orbits; and (\textit{iv}) stars with halo-like orbits. The presence of four distinctly different dynamical groups among the GCD sample suggests that there is not a single common origin for these stars and their origins are not like those of typical MW populations. We conclude that, if metal-rich GCs are responsible for these chemically unique stars, then there should be several groups of GCs contributing to the GCD population, e.g., those formed in-situ and those with an extragalactic origin \citep[see, e.g.,][]{Massari2019}.

\section{Conclusions} 
  
  We report on the discovery of a large population of relatively high-metallicity ($-0.7<$ [Fe/H] $<-0.1$) Globular Cluster Debris (GCD) stars identified in the interim data from the seventeenth data release of the APOGEE-2 Survey. 
  
The newly identified GCD stars are strongly enriched in nitrogen ([N/Fe] $\gtrsim+0.75$), and chemically distinct from MW and dwarf galaxy stars in almost all the chemical species examined so far, but with chemical patterns similar to those found in GC populations of comparable metallicity. We find that most of the relatively high-metallicity GCD stars lie on bulge- (stars with P-R orbits), disk- and halo-like orbits, and are likely part of the cumulative effect of many events, including partially or completely dissolved disk/bulge GCs of relative high metallicity, and/or dissolved GCs stars ejected from massive accretion events, such as the \textit{Gaia-}Enceladus-Sausage. Stars in our sample were mostly identified within the bulge, disk (likely the thick disk) and halo of the MW.
  
  The presence of a semi-regular variable (likely an AGB star) in our sample makes it possible that there is some contamination in our sample by variable stars and/or mass-transfer events, which could explain part of the chemically anomalous patterns at relatively high-metallicity. Future, long-term radial-velocity monitoring of our sample would naturally be the best course to establish the number of such sources formed through the binary or pulsating-star channels. 
  
  Finally, we conclude that, whatever process is responsible for the origin of the GCD stars, it is similar to that associated with the production of the unusual chemical-abundance patterns in relatively high-metallicity GCs.

  \section{Data Availability}  
  
The observational data underlying this article are from \textit{Gaia} EDR3, and from the 17th data release of the Sloan Digital Sky Survey (SDSS-IV), which are proprietary and will later be made publicly available. A short list of the APOGEE$-$Ids of all the potential members of the High-Metallicity Globular Cluster Debris can be found in the Table \ref{Table1}.

 \begin{table}
\begin{center}
\setlength{\tabcolsep}{3.0mm}  
\caption{APOGEE$-$Ids of the High-Metallicity Globular Cluster Debris Stars}
\begin{tabular}{cc}
\hline
APOGEE$-$Ids	&	APOGEE$-$Ids	\\ 
\hline
2M17372753$-$0425598 & 2M06410076$-$6926199 \\
2M05254390$-$0237519 & 2M06202897$-$6558047 \\ 
2M05495421$+$0043416 & 2M07084597$-$6227180 \\
2M11173689$+$0645217 & 2M05120630$-$5913438 \\
2M22534888$+$0919004 & 2M06040390$-$5600065 \\
2M08280055$+$1036106 & 2M08234846$-$4918149 \\
2M04392467$+$2348469 & 2M17571419$-$3328194 \\
2M16051430$+$2801104 & 2M08120643$-$3320409 \\
2M19261583$+$3653131 & 2M17255366$-$3208304 \\
2M02073982$+$3707297 & 2M18022530$-$2928338 \\
2M10392171$+$3828495 & 2M17480068$-$2922069 \\
2M05294163$+$3942524 & 2M17433446$-$2842533 \\
2M19233926$+$4003386 & 2M17372741$-$2816560 \\
2M19442885$+$4354544 & 2M17173130$-$2728590 \\
2M06383118$+$4646336 & 2M17305251$-$2651528 \\
2M23065670$+$4724235 & 2M16485601$-$2522202 \\
2M21070345$+$4733270 & 2M18341668$-$2333440 \\
2M22050390$+$5500110 & 2M17441535$-$2227514 \\
2M08513911$+$5550206 & 2M06301992$-$1604284 \\
2M12595431$+$5811394 & 2M11014631$-$1051279 \\
2M03521331$+$6755389 & 2M18531386$-$0942286 \\
\hline 
\end{tabular}  \label{Table1}
\end{center}
\end{table}    
  
\acknowledgments
We thank the anonymous referee for helpful comments that greatly improved the paper. T.C.B. acknowledges partial support for this work from grant PHY 14-30152: Physics Frontier Center / JINA Center for the Evolution of the Elements (JINA-CEE), awarded by the US National Science Foundation. D.M. is supported by the BASAL Center for Astrophysics and Associated Technologies (CATA) through grant AFB 170002, and by project FONDECYT Regular No. 1170121. B.B. acknowledge partial financial support from FAPESP, CNPq, and CAPES - Finance Code 001. L.C.V acknowledges the support of the postdoctoral Fellowship of DGAPA-UNAM, M\'exico, and the Fondo Nacional de Financiamiento para la Ciencia, La Tecnolog\'ia y la innovaci\'on ``FRANCISCO JOS\'E DE CALDAS", MINCIENCIAS, and the VIIS for the economic support of this research. 
\newline
Funding for the Sloan Digital Sky Survey IV has been provided by the Alfred P. Sloan Foundation, the U.S. Department of Energy Office of Science, and the Participating Institutions. SDSS-IV acknowledges support and resources from the Center for High-Performance Computing at the University of Utah. The SDSS website is www.sdss.org.
\newline
SDSS-IV is managed by the Astrophysical Research Consortium for the Participating Institutions of the SDSS Collaboration including the Brazilian Participation Group, the Carnegie Institution for Science, Carnegie Mellon University, the Chilean Participation Group, the French Participation Group, Harvard-Smithsonian Center for Astrophysics, Instituto de Astrof\'{i}sica de Canarias, The Johns Hopkins University, Kavli Institute for the Physics and Mathematics of the Universe (IPMU) / University of Tokyo, Lawrence Berkeley National Laboratory, Leibniz Institut f\"{u}r Astrophysik Potsdam (AIP), Max-Planck-Institut f\"{u}r Astronomie (MPIA Heidelberg), Max-Planck-Institut f\"{u}r Astrophysik (MPA Garching), Max-Planck-Institut f\"{u}r Extraterrestrische Physik (MPE), National Astronomical Observatory of China, New Mexico State University, New York University, the University of Notre Dame, Observat\'{o}rio Nacional / MCTI, The Ohio State University, Pennsylvania State University, Shanghai Astronomical Observatory, United Kingdom Participation Group, Universidad Nacional Aut\'{o}noma de M\'{e}xico, University of Arizona, University of Colorado Boulder, University of Oxford, University of Portsmouth, University of Utah, University of Virginia, University of Washington, University of Wisconsin, Vanderbilt University, and Yale University.
\newline
This work has made use of data from the European Space Agency (ESA) mission \textit{Gaia} (\url{http://www.cosmos.esa.int/gaia}), processed by the \textit{Gaia} Data Processing and Analysis Consortium (DPAC, \url{http://www.cosmos.esa.int/web/gaia/dpac/consortium}). Funding for the DPAC has been provided by national institutions, in particular the institutions participating in the \textit{Gaia} Multilateral Agreement.	
\newline
Simulations have been executed on HPC resources on the Cluster Supercomputer Atocatl from Universidad Nacional Aut\'onoma de M\'exico (UNAM). The Geryon2 cluster housed at the Centro de Astro-Ingenier\'ia UC was used for the calculations performed in this paper. The BASAL PFB-06 CATA, Anillo ACT-86, FONDEQUIP AIC-57, and QUIMAL 130008 provided funding for several improvements to the Geryon/Geryon2 cluster.


\end{document}